\documentclass[referee]{raa}     
\usepackage{graphicx,times}      
\usepackage{natbib}
\usepackage{amssymb,amsmath}
\bibpunct{(}{)}{;}{a}{}{,}

\usepackage[a4paper=true,pagebackref=true]{hyperref}
\hypersetup{colorlinks = true, linkcolor = green, anchorcolor = red, citecolor = blue, filecolor = red, pagecolor = red, urlcolor = red}

\DeclareGraphicsExtensions{.ps,.eps.gz,.epsi}

\newcommand{\teff}{\mbox{$T_{\rm eff}$}} \newcommand{\logg}{{\rm{log}~$g$}}
\newcommand{\feh}{{\rm [Fe/H]}} 
\newcommand{\ebv}{$E(B-V)$}

\begin{document}

   \title{Stellar loci IV. Red giant stars}

   \volnopage{Vol.0 (20xx) No.0, 000--000}      
   \setcounter{page}{1}          

   \author{Ruoyi Zhang
      \inst{1}
   \and Haibo Yuan
      \inst{1}
   \and Xiaowei Liu
      \inst{2}
   \and Maosheng Xiang
      \inst{3}
   \and Yang Huang
      \inst{2}
   \and Bingqiu Chen
      \inst{2}
   }

   \institute{Department of Astronomy, Beijing Normal University, Beijing 100875, China; {\it yuanhb@bnu.edu.cn}\\
        \and
             South-Western Institute for Astronomy Research, Yunnan University, Kunming 650500, China;\\
        \and
             Max-Planck Institute for Astronomy, Königstuhl, D-69117, Heidelberg, Germany\\
\vs\no
   {\small Received~~20xx month day; accepted~~20xx~~month day}}

\abstract{ In the fourth paper of this series, we present the metallicity-dependent Sloan Digital Sky Survey (SDSS) 
stellar color loci of red giant stars, 
using a spectroscopic sample of red giants in the SDSS Stripe 82 region. 
The stars span a range of 0.55 -- 1.2\,mag in color $g - i$,  $-$0.3 -- $-2.5$ in metallicity \feh,
and have values of surface gravity \logg~smaller than 3.5 dex.  
As in the case of main-sequence (MS) stars, the intrinsic widths of loci of red giants are also found 
to be quite narrow, a few mmag at maximum. 
There are however systematic differences between the metallicity-dependent stellar loci of red giants and MS stars.
The colors of red giants are less sensitive to metallicity than those of MS stars. 
With good photometry, photometric metallicities of red giants can be 
reliably determined by fitting the $u-g$, $g-r$, $r-i$, and $i-z$ colors simultaneously 
to an accuracy of 0.2 -- 0.25\,dex, 
comparable to the precision achievable with low-resolution spectroscopy for a signal-to-noise ratio of 10. 
By comparing fitting results to the stellar loci of red giants and MS stars, 
we propose a new technique to discriminate between red giants and MS stars based on the SDSS photometry.
The technique achieves completeness of $\sim$ 70 per cent and efficiency of $\sim$ 80 per cent 
in selecting metal-poor red giant stars of \feh~$\le -$1.2. 
It thus provides an important tool to probe the structure and assemblage history of the Galactic halo using red giant stars. 
}

   \authorrunning{Zhang et al.}            
   \titlerunning{Stellar loci IV. Red giant stars}  

   \maketitle

%
%
\section{Introduction}           
\label{sect:intro}

Red giants are intrinsically luminous objects,  
thus excellent tracers to probe the Galaxy, especially the Galactic halo.
However, it is a challenging task to select red giants based on optical photometry alone, due to the 
severe contamination from the foreground red dwarfs.

A number of earlier surveys of red giants select the targets based on the Mg\,I\,$b$ triplet and MgH band  
around 5,200\AA, detected in low-resolution objective prism spectra \citep{1985ApJ...291..260R,1990AJ....100.1181F},
or by intermediate-band photometry \citep{2000AJ....119.2254M}. The feature is more prominent in dwarfs than 
in giants, but also in metal-rich stars than in metal-poor stars (see Figure\,3 of \citealt{2000AJ....119.2254M}).
The {\rm Sloan Extension for Galactic Understanding and Exploration} ({\rm SEGUE}; \citealt{2009AJ....137.4377Y}) 
selects candidates of red giants on the basis that they are 
metal-poor in the halo. However, the efficiency of selection is quite low (e.g., for g = 17--18, 0.5 $\textless$ (g - r)$_0$ $\textless$ 0.6, their success rate is 45\%; for g = 17--18, 0.6 $\textless$ (g - r)$_0$ $\textless$ 0.8, the success rate decreases to only 28\%).

Stellar colors depend mainly on the effective temperature, 
but also to a modest degree on the metallicity and surface gravity, in particular the blue colors. 
With precise color measurements from modern digital sky surveys, such as the Sloan Digital Sky Survey (SDSS; \citealt{2000AJ....120.1579Y}), 
the Two Micro All-Sky Survey (2MASS; \citealt{2006AJ....131.1163S}), and the Wide-field Infrared Survey Explorer (WISE; \citealt{2010AJ....140.1868W}), 
and basic atmospheric parameters (effective temperature \teff, surface gravity \logg, and 
metallicity \feh) available from large scale stellar spectroscopic surveys, 
such as the SEGUE \citep{2009AJ....137.4377Y} and Large Sky Area Multi-Object Fiber Spectroscopy Telescope (LAMOST) Galactic surveys \citep{2012RAA....12..735D,2014IAUS..298..310L},
for large numbers of stars of various types, we are now in a position to (re-)investigate the dependence of stellar colors on the basic stellar parameters in unprecedented precision and quantitatively.

The repeatedly scanned equatorial Stripe 82  ($|{\rm Dec}| < 1.266\degr$, 20$^h$34$^m$ $< {\rm Ra} <$ 4$^h$00$^m$) 
of SDSS has delivered accurate photometry internally consistent at the 1 per cent level 
for about one million stars in $u,g,r,i,z$ bands \citep{2007AJ....134..973I},
as well as precise stellar parameters for over 40,000 stars with the SDSS Data Release 9 (DR9; \citealt{2012ApJS..203...21A}).
Using about 20,000 spectroscopically observed stars in the region as color standards, 
\cite{2015ApJ...799..133Y} have further re-calibrated the Stripe 82 with the stellar color regression (SCR) method,
achieving an unprecedented internal accuracy of
about 0.005, 0.003, 0.002, and 0.002\,mag in colors $u-g$, $g-r$, $r-i$, and $i-z$, respectively.
By combining the re-calibrated photometric data and the spectroscopic parameters of SDSS Stripe 82,
\citet[hereafter Paper\,I]{2015ApJ...799..134Y} build a large, clean sample of main sequence (MS) stars with accurate colors (about 1 per cent)
and well-determined metallicities (about 0.1\,dex) to investigate the metallicity dependence and intrinsic widths of SDSS stellar color loci.
They demonstrate that the intrinsic widths of loci of MS stars, after being corrected for the effects of metallicity, are only a few mmag at most.
They also find that outliers of the metallicity-dependent stellar loci are mainly contributed by binaries. 
\citet[hereafter Paper\,II]{2015ApJ...799..135Y} proposes a Stellar Locus OuTlier (SLOT) method to 
estimate the binary fraction for field stars and find interesting trends of the fraction with stellar colors and metallicity.
\citet[hereafter Paper\,III]{2015ApJ...803...13Y} further use the metallicity-dependent stellar loci to 
determine state-of-the-art photometric metallicities for about 0.5 million FGK stars in the Stripe\,82 region. 
A precision of 0.1 -- 0.2\,dex is achieved for most of the stars. 

In this paper, we focus on red giant stars.
The paper is organized as follows. 
In Section 2, we introduce the spectroscopic and photometric data used.
In Section\,3, we investigate the metallicity dependence and intrinsic widths of stellar loci of red giant stars, following Paper\,I.
The photometric metallicities of red giants are presented and examined in Section\,4, following Paper\,III.
In Section\,5, we propose and test a new technique to discriminate red giant stars from red dwarfs based on the SDSS photometry. 
The summary is given in Section\,6. 


\section{Data}
\label{sect:Data}

Following Paper\,I, we first select stars from the SDSS DR9 in the Stripe 82 region 
that have been observed spectroscopically with  
a spectral signal-to-noise ratio S/N $>$ 10 and are listed 
in the re-calibrated photometric catalogs of Stripe 82 \citep{2007AJ....134..973I,2015ApJ...799..133Y}
This yields 34,906 stars. The stars are then dereddened using the reddening values given by the dust map of 
\cite[hereafter SFD]{1998ApJ...500..525S}. The empirical reddening coefficients of \cite{2015ApJ...799..133Y}, 
derived using a star pair technique \citep{2013MNRAS.430.2188Y}, are used.
Although the SFD map suffers some calibration issue that it may over-estimate reddening, such effect can be corrected for by using our empirical reddening coefficients.
Finally, stars of a line-of-sight extinction \ebv~$\le$ 0.15\,mag, surface gravity \logg~$\le$ 3.5\,dex, 
dereddened color\footnote{All colors and magnitudes quoted in the current paper referred to dereddened values unless specified otherwise.}
$0.55 \le g-i \le 1.2$\,mag, effective temperature \teff~$\ge$ 4,300\,K, and metallicity $-2.5 \le \feh~\le -0.3$, are selected.
Here the basic stellar parameters, effective temperature, surface gravity, and metallicity, 
are determined with the SEGUE Stellar Parameter Pipeline (SSPP; \citealt{2008AJ....136.2022L,2008AJ....136.2050L}).
The final red giant sample contains 2,252 stars.
The distribution of the stars in the $g-i$ and \feh~plane is shown in the top left panel of Fig.\,1. 
Compared to 24,492 MS sample stars of Paper\,I, the number of red giants of the current sample is much smaller.
Their metallicities extend however to \feh~=$-$2.5, representing some of the very metal-poor red giants in the Galactic halo.
The photometric errors as a function of the observed magnitudes before reddening corrections are also shown in Fig.\,1.
The behaviors are very similar to those of the MS sample of Paper\,I.
For $u$ band, the photon counting noises start to dominate the errors at 
$u \ga 19$\,mag. The error are about 0.01 mag at $u=19.0$\,mag and 0.05 mag at $u=21.5$\,mag.
For $g$, $r$, and $i$ bands, the errors are dominated by the calibration uncertainties and thus essentially 
constant, at a level of $0.006\pm0.001$, $0.005\pm0.001$, and $0.005\pm0.001$ mag, respectively, for the three bands.
For $z$ band, the photon counting noises dominate at $z \ga 18$\,mag, and the errors
are about 0.01 mag at $z=18.2$\,mag and 0.02 mag at $z=19.0$\,mag.

\begin{figure*}
   \centering
    \includegraphics[width=8cm]{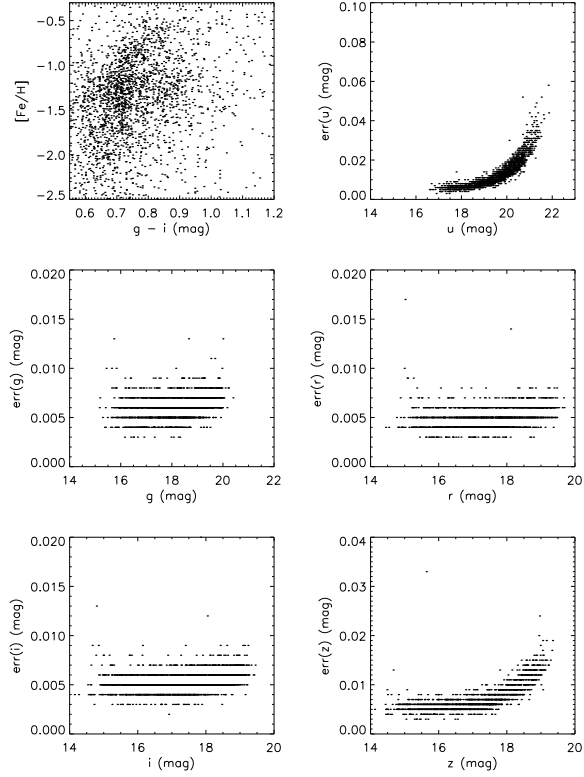}
   \caption{
   {\em Top left panel:}\, Values of [Fe/H] plotted against color $g-i$
   for the selected red giant sample. {\em The remaining panels:}\, Photometric errors as a function of the observed magnitudes before reddening corrections. 
   }
   \label{}
   \end{figure*}

\section{Metallicity-Dependent Stellar Loci}
Using the sample above, we have carried out a global two-dimensional polynomial fit to colors $u-g$, $g-r$, $r-i$, and $i-z$ 
as a function of color $g-i$ and metallicity \feh. 
Again we have neglected possible effects of \logg~on the colors.
In all cases, a 3rd-order polynomial of 10 free parameters is used.
Two-sigma clipping is performed to reject outliers when fitting the data. 
The resultant fit coefficients are listed in Table\,1.
Note the sums of the corresponding coefficients for colors $g-r$ and $r-i$ are exactly zero or one.
The fit residuals as a function of $g-i$, \feh~and \logg~ are shown in Fig.\,2. 
The median values and standard deviations are delineated by red lines. 
Note that the fit residuals in color $u-g$ show a weak systematic variation with \logg.

Fig.\,3 compares the stellar loci of MS stars and red giants at 
different metallicities ranging from $-2.5$ -- 0.0.
Note that in this Figure the stellar loci of MS stars at \feh~= $-2.5$ and those of red giants at \feh~= 0.
are calculated by extrapolating the fits, and might suffer some systematic errors. However, considering that metallicities of our giant stars are up to $-$0.3, those of the MS stars are down to $-$2.0, and the smooth variations 
of the loci with respect to \feh~, the systematic errors should be very small.
Similar to MS stars, $u-g$ is the color most sensitive to metallicity for red giants as well. 
Colors $g-r$, $r-i$, and $i-z$ also show some modest sensitivity to metallicity. 
The variations are larger in metal-rich stars than in metal-poor ones. 
However, in general, the variations of colors in giants as \feh~varies are much lower than in MS stars.
At \feh~= $-$1.0, one dex decrease in \feh~ leads to only modest 0.2 and small 0.017 mag decrease in colors $u-g$ and $g-r$, respectively,
and only marginal 0.01 mag increase in color $i-z$. 
For the same $g-i$ color and metallicity \feh, the predicted colors $u-g$, $g-r$, $r-i$, and $i-z$ of giants and dwarfs 
differ significantly, particularly in $u-g$.
For instance, at \feh~= $-$2.0 and $g-i$ = 1.0, the color differences between giants and dwarfs are 0.10 for $u-g$, and about 0.01 in colors $g-r$, $r-i$, and $i-z$.
Giants of a given metallicity show colors very similar to MS stars but with lower metallicities 
(see \citet[Figure\,6]{2015ApJ...803...13Y} for a qualitative comparison).
  
To investigate the intrinsic widths of stellar loci of red giant stars, we divide the sample into 
bins of color and metallicity. For each bin, a Gaussian is used to fit the distribution of fit residuals.
To minimize the effects of photometric errors,
only stars of errors smaller than 0.01 mag in $u$ are used in the case of color $u-g$,
and likewise in $z$ in the case of color $i-z$.  
The results for two typical color bins (centered at $g-i$ = 0.7 and 1.0\,mag) and two metallicity 
bins (centered at $\feh = -0.8$ and $-1.5$) are plotted in Fig.\,4.  

\begin{figure*}
\centering
\includegraphics[width=\textwidth]{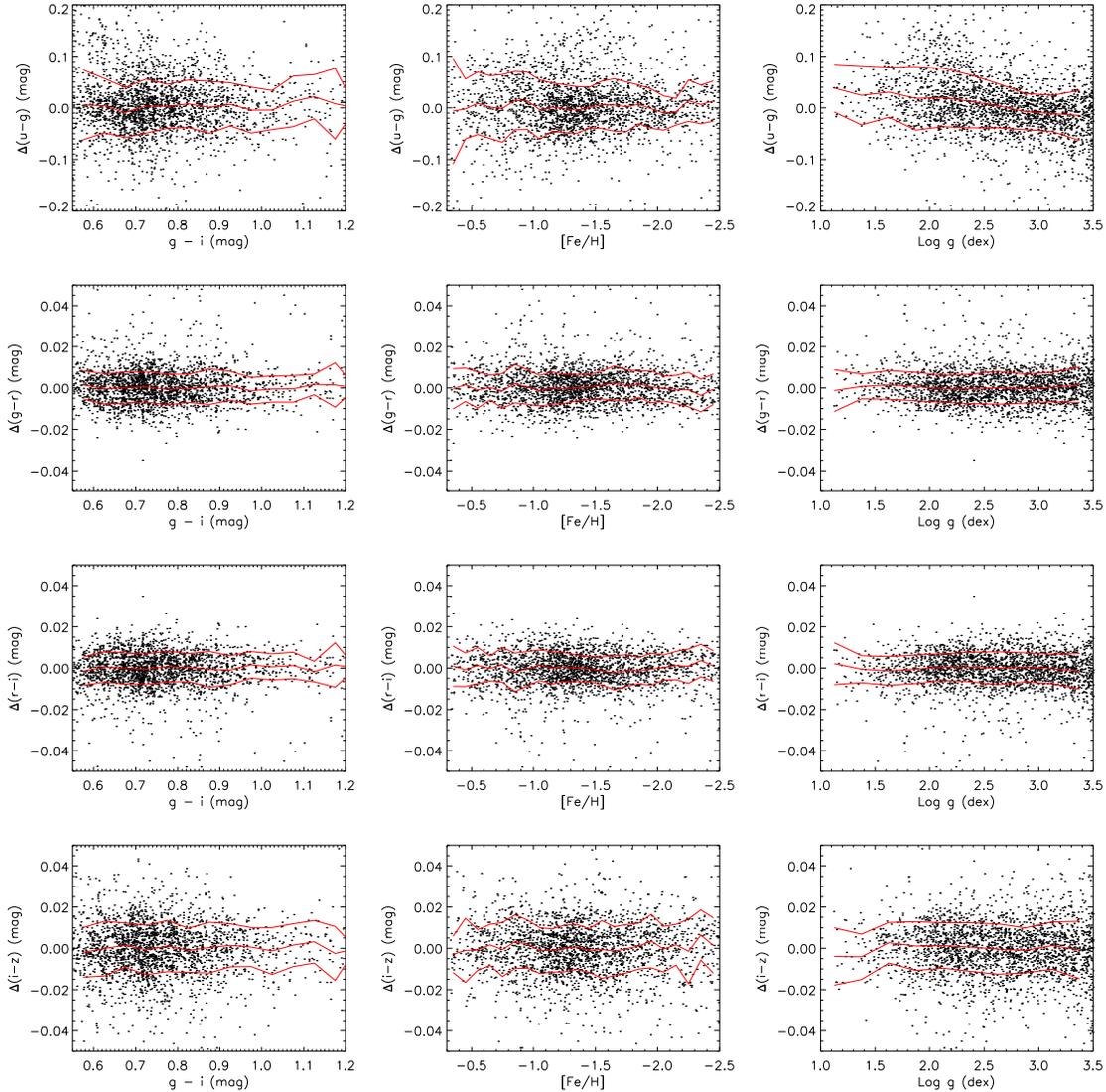}
\caption{
Fit residuals as a function of color $g-i$ (left),  metallicity \feh~(middle), 
and surface gravity \logg~(right).
Lines delineating the median and standard deviation of the residuals are over-plotted in red. 
}
\label{}
\end{figure*}

\begin{figure*}
\centering
\includegraphics[width=\textwidth]{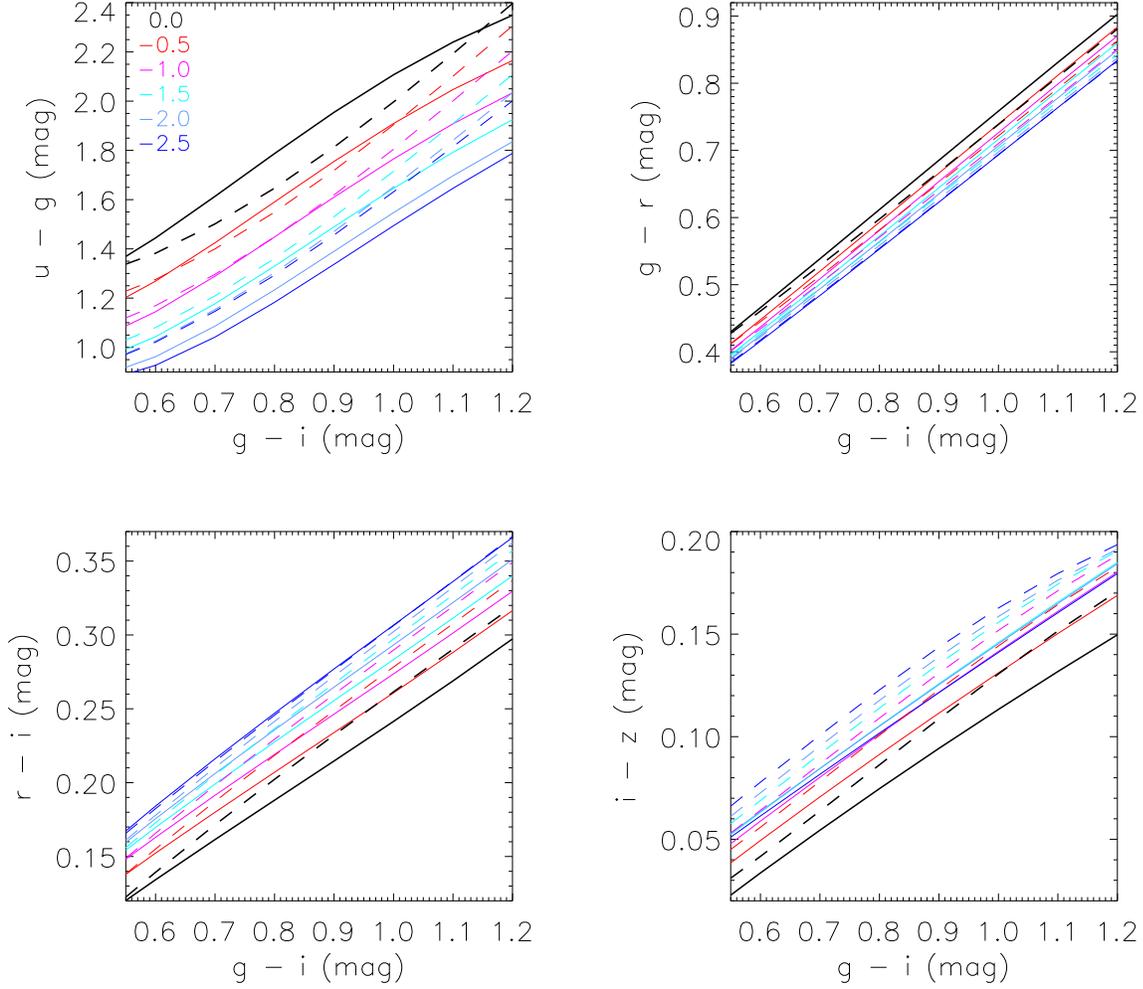}
\caption{
Stellar loci of MS stars (solid lines) and red giants (dashed lines) of different metallicities 
of ranging from $-2.5$ to 0.0.
}
\label{}
\end{figure*}

\begin{table*} 
\centering
\caption{Fit Coefficients.}
\label{}
\begin{tabular}{lrrrr} \hline\hline
Coeff. & $u-g^a$ & $g-r^a$ & $r-i^a$ & $i-z^a$  \\
$a_{0}$ &   1.4630 &   0.0957 &  $-$0.0957 &  $-$0.0551 \\
$a_{1}$ &   0.3132 &   0.0370 &  $-$0.0370 &  $-$0.0229 \\
$a_{2}$ &  $-$0.0105 &   0.0120 &  $-$0.0120 &  $-$0.0165 \\
$a_{3}$ &  $-$0.0224 &   0.0020 &  $-$0.0020 &  $-$0.0033 \\
$a_{4}$ &  $-$1.5851 &   0.5272 &   0.4728 &   0.0762 \\
$a_{5}$ &  $-$0.2423 &  $-$0.0026 &   0.0026 &  $-$0.0365 \\
$a_{6}$ &  $-$0.0372 &   0.0019 &  $-$0.0019 &  $-$0.0006 \\
$a_{7}$ &   2.8655 &   0.1645 &  $-$0.1645 &   0.1899 \\
$a_{8}$ &   0.0958 &   0.0057 &  $-$0.0057 &   0.0244 \\
$a_{9}$ &  $-$0.7469 &  $-$0.0488 &   0.0488 &  $-$0.0805 \\
\hline
\end{tabular}
\begin{description}
\item[$^a$]  $f(x,y)=a_0+a_1y+a_2y^2+a_3y^3+a_4x+a_5xy+a_6xy^2+ $ \\ 
             $a_7x^2+a_8yx^2+a_9x^3$ , where $x$ $\equiv$ $g-i$ and $y$ $\equiv$ \feh.
\end{description}
\end{table*}

The average photometric errors are $0.008\pm0.002$
$0.006\pm0.001$, $0.005\pm0.001$, $0.005\pm0.001$, and $0.006\pm0.001$\,mag in $u, g, r, i$, and $z$ bands, respectively,
very close to those found for MS stars in Paper\,I. 
The color calibration uncertainties are about 0.005, 0.003, 0.002, and 0.002\,mag for colors $u-g$, $g-r$, $r-i$, and $i-z$, 
respectively \citep{2015ApJ...799..133Y}.
The typical uncertainties of \feh~yielded by the SSPP is about 0.13 dex \citep{2008AJ....136.2050L}. 
Given the above uncertainties, one expects dispersion of 
0.028, 0.0086, 0.0078, and 0.0085 mag in colors $u-g$, $g-r$, $r-i$, and $i-z$, respectively.
Here the dispersion are computed using the polynomial equation in the footnote of Table\,1 and
the corresponding coefficients, along with previously known errors in the photometry, calibration, and spectroscopic metallicity.
The dispersion yielded by the above Gaussian fits to the residuals of the whole selected sample are 
0.031, 0.0071, 0.0071, and 0.0105\,mag, respectively, very close to the expected values, 
suggesting that, similar to what we find in Paper~I for MS dwarfs,  
the intrinsic widths of SDSS stellar loci of red giant stars, similar to those of MS stars, are also very narrow 
and at maximum a few mmag.
Unlike in the case of MS stars, the residual distributions are well fitted by Gaussian function, suggesting that binaries composed of two red giants are rare.
The large dispersion in color $u-g$ is mainly contributed by the uncertainties in \feh~and by some unaccounted for dependence of the 
color on \logg.
The dispersion in other colors are dominated by photometric errors.
The dispersion show very small variations among individual bins of color and metallicity. 
As in Paper\,I, in the above analysis, we have neglected the possible effects of variations in the [$\alpha$/Fe] abundance ratio, 
of variable stars, and uncertainties in the reddening corrections.
As argued in Paper\,I and also shown in Paper\,II, these effects are ignorable.

As a further test of the above results,  we examine the intrinsic width of stellar loci of red giant branch stars in four globular 
clusters (GCs) M\,92, M\,13, M\,3, and M\,5. For each GC, we select a sample of red giant branch stars in the color range 
of $0.55 < g-i < 1.2$\,mag from the photometric catalogs of \cite{2008ApJS..179..326A}, 
deduced with the DAOPHOT/ALLFRAME \citep{1987PASP...99..191S,1994PASP..106..250S} suite of programs for crowded field photometry.
Note that the standard SDSS photometric pipelines \citep{2002SPIE.4836..350L} can not deal with crowded fields properly.
To remove stars that have poor values of goodness-of-fit, 
we filter the data based on the sharpness and $\chi$ values from DAOPHOT, excluding stars 
of $|sharpness| > 1$ or $\chi > 1.5 + 4.5 \times 10^{-0.4 \times (m - m_{\rm 0})}$, where $m_{\rm 0}$ = 15.5\,mag for $u$, 
16.0\,mag in $gri$, and 15.0\,mag in $z$ \citep{2008ApJS..179..326A}.
The remaining stars are reddening corrected using the SFD extinction map and the reddening coefficients of \cite{2013MNRAS.430.2188Y}.
Their $u-g$, $g-r$, $r-i$, and $i-z$ colors are fitted as a function of $g-i$ color using third-order polynomials.
The histogram distributions of fit residuals are shown in Fig.\,5.
The dispersion of fit residuals range between 0.032 -- 0.06, 0.009 -- 0.015, 
0.009 -- 0.015, and 0.016 -- 0.022\,mag in colors $u-g$, $g-r$, $r-i$, and $i-z$, respectively.
The random errors of the DAOPHOT photometry of \cite{2008ApJS..179..326A} are about 0.02\,mag in $u$ and 0.01\,mag in $griz$ bands 
at the bright ends. 
The dispersion are thus comparable to the photometric errors, 
consistent with the fact that the intrinsic widths of SDSS stellar color loci of red giants must be very small, at maximum a few mmag.
The relatively large dispersion seen in $u-g$ color is likely caused by the calibration errors, or 
the presence of multiple populations in those GCs \citep{2011A&A...525A.114L}, or both.

\begin{figure*}
\centering
\includegraphics[width=\textwidth]{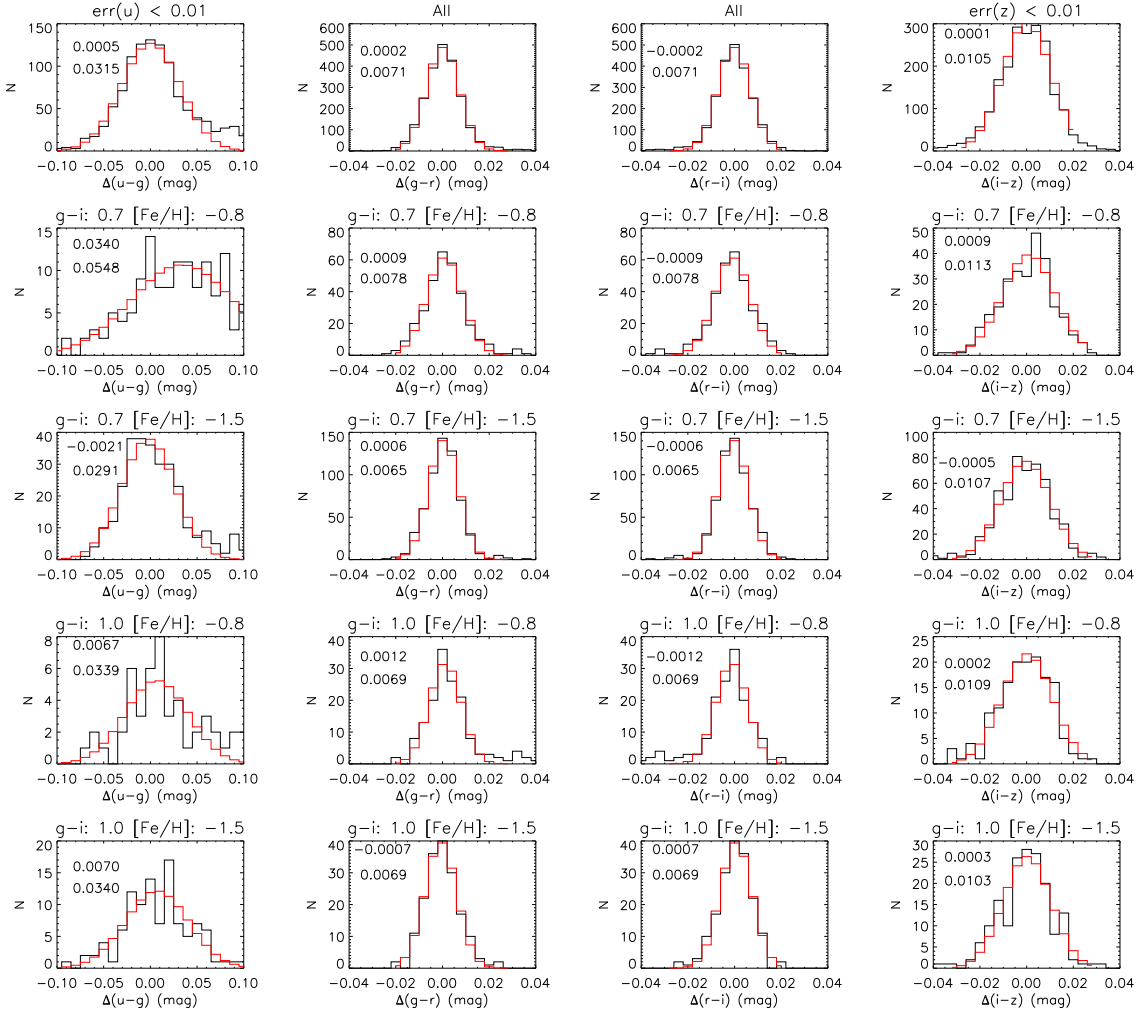}
\caption{
Histograms of fit residuals for 
all the sample stars (of $u$ band photometric errors less than 0.01\,mag in the case $u - g$ and 
of $z$ band photometric errors less than 0.01\,mag in the case of $i - z$), and for some selected bins of color
$g-i$ and metallicity \feh. 
The central values of $g-i$ and \feh~of the bin are labeled at the top of each panel. 
The bin widths are 0.4 mag in $g - i$ and 0.4 in \feh, respectively. 
Over-plotted in red are Gaussian fits to the distributions.
The mean and dispersion of the fitted Gaussian are labeled. 
}
\label{}
\end{figure*}

\begin{figure*}
\centering
\includegraphics[width=\textwidth]{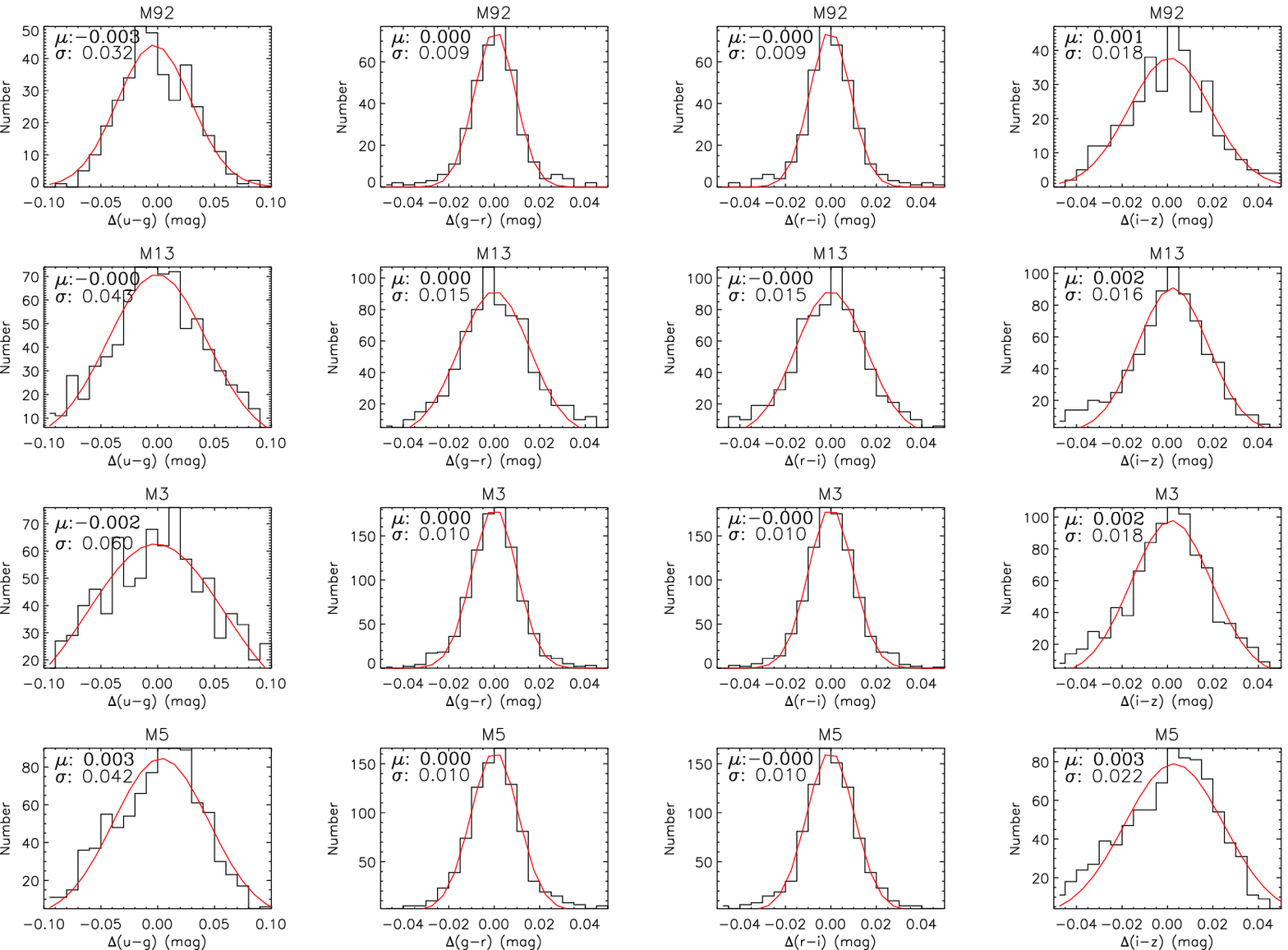}
\caption{
Histograms of fit residuals for
selected samples of red giant branch stars of GCs M\,92, M\,13, M\,3, and M\,5.
The name of each GC is marked at the top of each panel.
Gaussian fits to the distributions are over-plotted in red.
The mean and dispersion of the Gaussian are labeled. 
}
\label{}
\end{figure*}

\subsection{Photometric Metallicity}
In Paper\,III, we develop a  minimum  $\chi^2$ technique to estimate the photometric metallicity of MS stars
by fitting the reddening corrected colors in $u-g$, $g-r$, $r-i$, and $i-z$ with the corresponding values predicted by the 
metallicity-dependent stellar loci presented in Paper\,I, for a given set of intrinsic $g-i$ color and metallicity \feh. 
The optimal intrinsic $g-i$ color and \feh, corresponding to the minimum  $\chi^2$ value, $\chi_{\rm min}^2$,
is obtained by a brute-force algorithm. The one-sigma error of \feh~is also estimated.
Here we use the same technique 
to estimate the photometric metallicities of our sample red giants.
For a given star, we vary \feh~value from $-2.5$ -- 0.0, stepping 0.01 at a time.

We first apply the technique to the sample of red giant stars of Stripe\,82 presented in the previous section. 
A few stars of \feh~richer than $-$0.3 are also included here.
The left panel of Fig.\,6 shows the histogram distribution of $\chi_{\rm min}^2$ values of the sample.
Only 13.5, 6.3, and 2.8 per cent stars show a $\chi_{\rm min}^2$ value larger than 3, 5, and 10, respectively. 
The right panel of Fig.\,6 compares the
photometric metallicities thus derived with the spectroscopic values of SDSS DR9.
The distribution of differences shows a negligible offset and has a dispersion of 0.27\,dex.

To understand what contributes to the above dispersion, 
1,344 duplicate observations of comparable spectral SNRs of stars that fall in the
parameter ranges of the selected giant sample are selected from the SDSS DR9
to estimate the random errors of \feh~yielded by the SSPP pipeline.
Following Paper\,II, considering the relatively narrow range of effective temperature of red giants,
the random errors of \feh, estimated from those duplicate observations, 
are fitted as a function of SNR and \feh~of the following form:
\begin{align}
\sigma_{\rm {ran}}({\rm [Fe/H]}) = \nonumber\\ 
& a_0 +  a_1 \times {\rm [Fe/H]} + a_2 \times ({\rm [Fe/H]})^2 +\nonumber\\ 
&a_3 \times {\rm SNR} + a_4 \times {\rm SNR} \times {\rm [Fe/H]} + \nonumber\\ 
&a_5 \times {\rm SNR}^2.
\end{align}
Only stars of SNRs between 10 -- 50 are used in the fitting.
The resultant fit coefficients $a_0$ -- $a_5$ are 0.28, $-$0.0040, 0.0018, $-$0.0096, $1.4\times 10^{-4}$, and $9.5\times 10^{-5}$, respectively.
For SNR = 10, the random errors of \feh~yielded by the SSPP 
are thus about 0.20, 0.20, and 0.21\,dex for \feh~= 0, $-$1, and $-$2, respectively.
For SNR = 50, the corresponding values are 0.04, 0.05, and 0.07\,dex for \feh~= 0, $-$1, and $-$2, respectively.
The random errors of the test sample range from 0.04 to 0.22\,dex, with a median value of 0.11\,dex.
These results suggest that the photometric metallicities for red giant stars have a typical error between 0.2 -- 0.25\,dex.
The median and mean uncertainties of photometric metallicities deduced by the method for the selected giant sample are
0.15 and 0.18\,dex, respectively, suggesting that the random errors of metallicity of those red giants 
may have been slightly underestimated.

The top panels of Fig.\,7 plot differences between the spectroscopic metallicities of SDSS DR9 and 
the photometric values estimated above for the sample as a function of \feh, $g-i$, and \logg, respectively.
No systematic dependence of the differences on \feh~or $g-i$ is found.
However, some weak dependence on \logg~is seen, clearly a consequence of the weak systematic dependence 
of the fit residuals in $u-g$ on \logg~(the top right panel of Fig.\,2).
The scatter increases slightly for stars of lower metallicities, bluer colors, and lower surface gravities. 
The bottom panels of Fig.\,7 plot the estimated uncertainties of photometric metallicity 
as a function of error of color $u-g$, \feh, and $g-i$.
A good linear correlation is seen between the uncertainties of \feh~and the errors of color $u-g$.
. A linear fit yields, 
\begin{equation}
  \sigma({\rm [Fe/H]}) = 7.8\times \sigma (u-g) + 0.03.
\end{equation}
Compared to MS stars, the uncertainties of photometric metallicities of red giant stars are more sensitive to the errors of color $u-g$, 
due to the fact that the $u-g$ colors of giants are less sensitive to \feh~ than for MS stars.
Similarly, the uncertainties of \feh~estimates are larger for metal-poor stars than for metal-rich stars.
The values are about 0.14\,dex at \feh~$>$ $-$1.7 and increase to 0.24\,dex at lower metallicities.
The metallicity uncertainties show only a weak dependence on color in the range considered in the current work.

\begin{figure*}
\centering
\includegraphics[width=8cm]{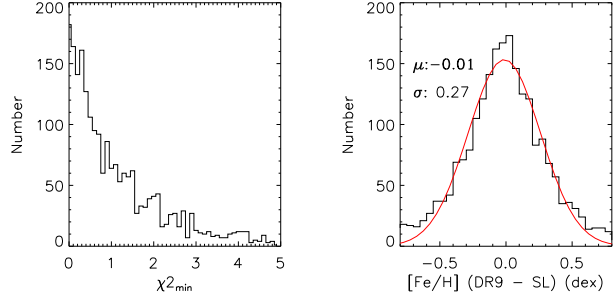}
\caption{
Histogram distributions of $\chi_{\rm min}^2$ values of the test sample (left) and the differences between the
photometric metallicities derived in the current work and spectroscopic values from SDSS DR9 (right).
The red curve is a Gaussian fit to the distribution, with the mean and dispersion marked.
}
\label{}
\end{figure*}

\begin{figure*}
\centering
\includegraphics[width=\textwidth]{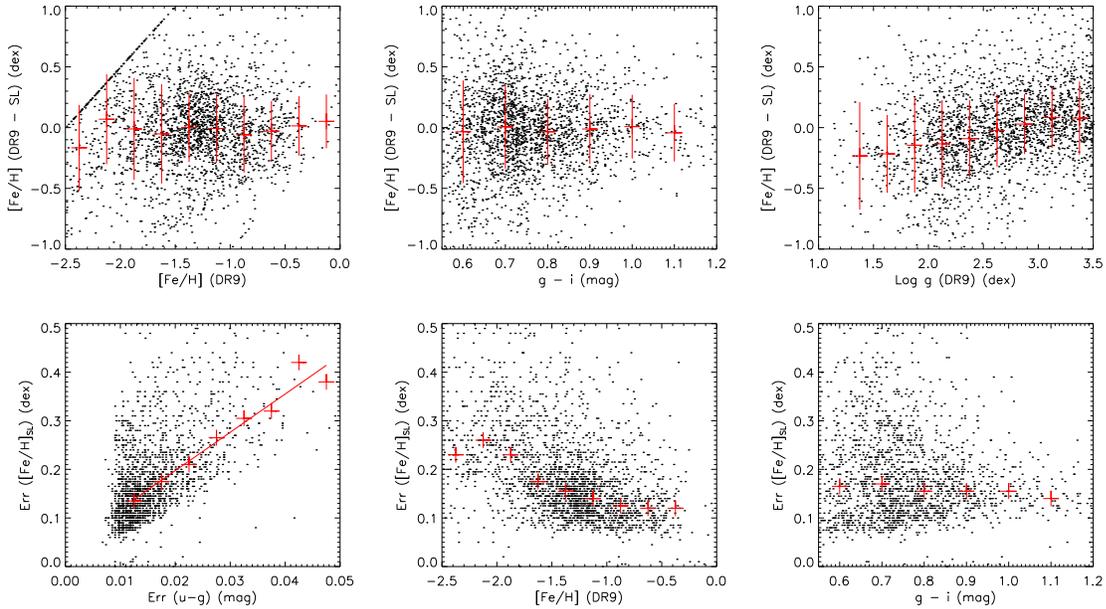}
\caption{
{\em Top panels:}\,Differences of spectroscopic and photometric metallicities plotted  
as a function of spectroscopic metallicity (left), $g-i$ color (middle) and \logg~(right).
The sample is divided into bins of \feh,  $g-i$, and \logg~of widths 0.25, 0.1\,mag, and 0.25\,dex, respectively.
For each bin, the median and dispersion values are over-plotted in red.
{\em Bottom panels:} Uncertainties of photometric \feh~of the test sample plotted against error in color $u-g$ (left), 
spectroscopic \feh~from the SDSS DR9 (middle), and color $g-i$ (right), respectively.
The sample is divided into bins of $\sigma (u-g)$, \feh, and $g-i$ of widths 0.005\,mag, 0.25\,dex, and 0.1\,mag, respectively.
For each bin, the median and dispersion values are over-plotted in red.
The red line in the left panel is a linear fit to the data.
}
\label{}
\end{figure*}

We have also tested the accuracy of photometric metallicities 
using the same sets of red giant stars in the four GCs M\,92, M\,13, M\,3, and M\,5 
selected in the previous Section.
Their distributions in the ($g-i$, $r$) color-magnitude diagram are shown in the first column of panels of Fig.\,8.
The second column of panels of Fig.\,8 plot photometric metallicities against $g-i$ colors.
For comparison, member red giants spectroscopically observed by the SDSS are also over-plotted in red in the 
first and  second columns.
The third and fourth columns of panels show histogram distributions of photometric 
and spectroscopic metallicities, as deduced here and from SDSS DR9, respectively. 
Over-plotted are Gaussian fits to the distributions, with the centers and dispersion of the fits labeled.
The photometric metallicities deduced for M\,92, M\,13, M\,3, and M\,5 show no 
obvious dependence on $g-i$ color,
and are $-1.51\pm0.26$, $-1.28\pm0.29$, $-1.28\pm0.37$, and $-0.94\pm0.23$, respectively.
The relatively small dispersion suggest that good photometry such as the SDSS can deliver reasonably precise estimates of 
metallicities of red giant stars.
Note that the larger dispersion of M\,3 is caused by the larger photometric errors of color $u-g$.

The mean photometric metallicities deduced here are found to be systematically higher than those 
from high-resolution spectroscopy as well as those from the SDSS DR9.
The metallicities of M\,92, M\,13, M\,3, and M\,5 from high-resolution spectroscopy are 
$-$2.38, $-$1.6, $-$1.5, and $-$1.26, respectively \citep{2004oee..sympE..33K}.
The spectroscopic metallicities of M\,92, M\,13, and M\,3 from the SDSS DR9 are 
$-2.32\pm0.08$, $-1.55\pm0.11$, and $-1.53\pm0.16$, respectively (M\,5 was not targeted by the SDSS).
The discrepancies between the photometric and spectroscopic metallicities are thus about 
0.9, 0.3, 0.2 and 0.3\,dex for M\,92, M\,13, M\,3, and M\,5, respectively.

The discrepancies can be caused by problems related to 1) The SDSS calibration of \feh~for giant stars; 
2) The values of \feh~from high resolution spectroscopy of \cite{2004oee..sympE..33K}; 
3) The reddening corrections; 
4) The metallicity-dependent stellar loci of the current work; and 
5) The color calibrations of SDSS photometry of the four GCs.
Given the good agreement between the metallicities given by the SDSS DR9, \cite{1996AJ....112.1487H}, and \cite{2004oee..sympE..33K} as well, 
Possibilities 1) and 2) are very unlikely.
The \ebv~values of the four GCs as given by the SFD extinction map are 0.02, 0.02, 0.04, and 0.01\,mag, respectively, 
too small to have a big effect on the photometric metallicities presented here. 
Given the small fit residuals when deriving the metallicity-dependent stellar loci, Possibility 4) is unlikely either.
We are now left with Possibility 5).
To account for the photometric and spectroscopic metallicity discrepancy of M\,92, 
one only needs a color calibration error of 0.045\,mag in $u-g$, 
or 0.06\,mag in $g-i$, or 0.025\,mag in $u-g$ and 0.03\,mag in $g-i$, or some other combinations. 
Similarly to account for the discrepancies of M\,13, M\,3, and M\,5, one only needs
a color calibration error of 0.03\,mag in $u-g$,
or 0.04\,mag in $g-i$, or 0.015\,mag in $u-g$ and 0.02\,mag in $g-i$, or some other combinations.
Here the calibration errors refer to those relative to the calibration of Stripe\,82, and could be 
caused by errors in the photometric zero point, flat-fielding, unaccounted fast variations 
of the atmospheric extinction, and/or non-linearity of the detectors. 
Corroborative evidence that the SDSS photometric calibration of the four clusters is to blame is presented in the 
next Section (c.f. Fig.\,12).
\cite{2013ApJ...763...65A} compare the SDSS photometry for several clusters from \cite{2008ApJS..179..326A}, including the four GCs used 
in the current work, based on the SDSS ubercal calibration \citep{2008ApJ...674.1217P}. 
The comparison reveals significant photometric zero-points of those clusters and the 
trends are consistent with what expected in order to reconcile the photometric and spectroscopic metallicity estimates, as described above. 
However, it is likely that a constant zero-point correction just serves as the zero order 
approximation. Future work is needed to investigate and correct for 
the calibration errors as functions of spatial position, magnitudes, and colors.

\begin{figure*}
\centering
\includegraphics[width=160mm]{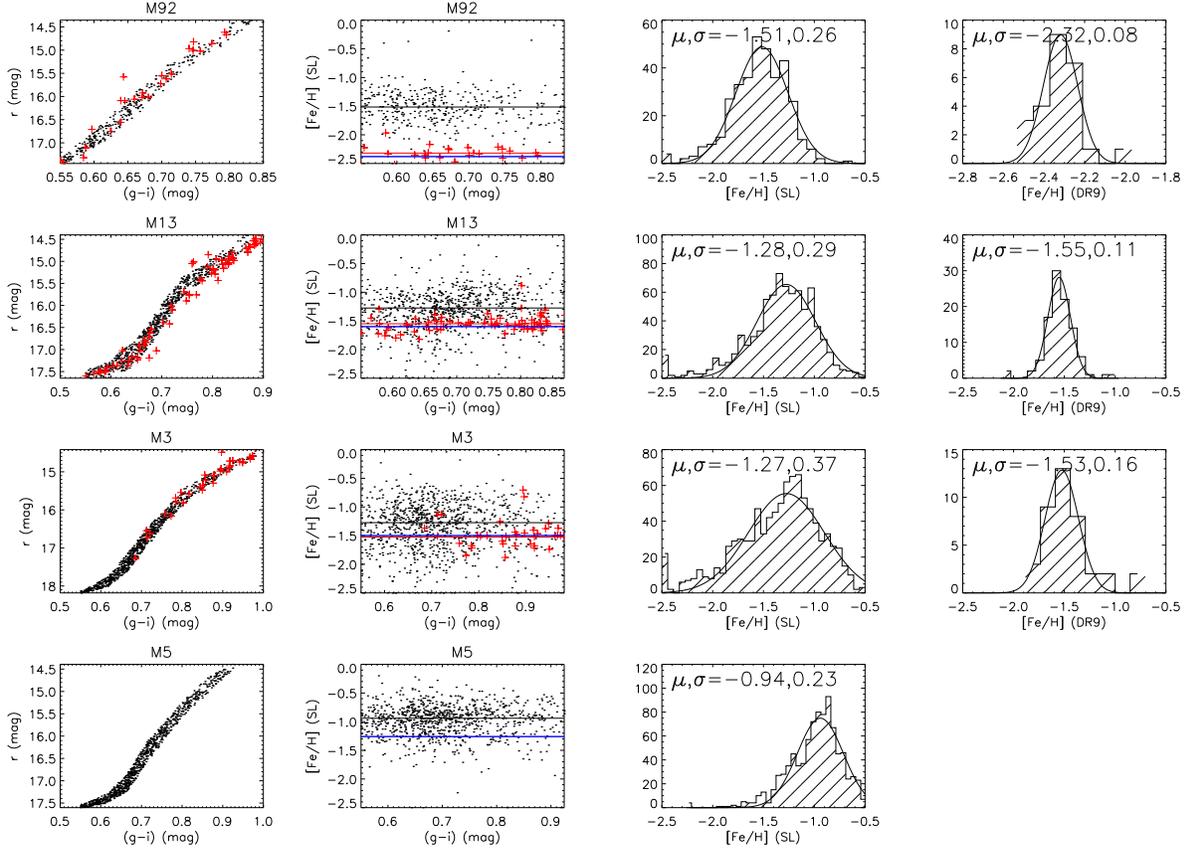}
\caption{
Photometric metallicities of red giant stars of GCs M\,92, M\,13, M\,3, and M\,5, and comparison with other estimates.
The name of the GC concerned is marked on the top of each panel.
{\em Column 1:}\, ($g-i$) versus $r$ color-magnitude diagram of the selected red giant stars in the four GCs.
Red pluses denote member stars spectroscopically targeted by the SDSS. 
{\em Column 2:}\, Photometric metallicities derived plotted against color $g-i$.
Red pluses denote member stars spectroscopically observed by the SDSS. 
The black, red, and blue lines
in panels of the second column denote the mean photometric metallicities, the mean spectroscopic metallicities from SDSS DR9,
and the average metallicities from high-resolution spectroscopy \citep{2004oee..sympE..33K}, respectively.
{\em Column 3:}\, Histogram distributions of photometric metallicities. 
{\em Column 4:}\, Histogram distributions of spectroscopic metallicities from SDSS DR9. 
In the 3rd and 4th columns, also over-plotted are Gaussian fits to the distributions,
and the central values and dispersion of the fits are labeled.
Note that no member stars of M\,5 have been spectroscopically targeted by the SDSS.
}
\label{}
\end{figure*}

\subsection{Selection of Red Giants}
There are systematic differences between the stellar loci of MS stars and red giant stars.
With this in mind, we propose in this Section a way to discriminate between red giants and red dwarfs using the stellar loci as a tool.
For a given target star, we determine its optimal value of \feh, assuming it is a dwarf or giant, 
denoted as \feh~(SL$_{\rm D}$) and \feh~(SL$_{\rm G}$), 
and the associated $\chi_{\rm min}^2$, denoted as $\chi_{\rm min}^2 (SL_D)$ and $\chi_{\rm min}^2 (SL_G)$, respectively. 
If $\chi_{\rm min}^2 (SL_D)$ is larger than  $\chi_{\rm min}^2 (SL_G)$, then the target star is likely a red giant and vice versa.

To test this idea, we select a sample of 16, 375 stars from the SDSS DR9 in the Stripe 82 region
that have been observed spectroscopically with a spectral SNR $>$ 10, and are listed
in the re-calibrated photometric catalogs of Stripe 82 \citep{2007AJ....134..973I,2015ApJ...799..133Y},
and having a $g-i$ color between 0.55 -- 1.2\,mag,  
an \feh~value between $-$2.5 -- 0.0, and an \ebv~value smaller than 0.15\,mag.
The sample includes 2,263 red giants (\logg~$<$ 3.5\,dex) and 14,112 MS stars (\logg~$>$ 3.5\,dex). 
The photometric metallicities of the sample stars are then estimated using the two sets of stellar loci,
one for giants and another for dwarfs, respectively.  
The results are presented in Fig.\,9.

The top panels of Fig.\,9 compare the photometric metallicities derived using the loci of either MS stars or red giants 
with the spectroscopic values from SDSS DR9.
The black and red dots denote MS stars and red giants in the sample, respectively. 
It is evident that when the correct set of stellar loci is used 
then the photometric metallicities agree well with the spectroscopic values. 
However, when the wrong set of stellar loci is used instead, the photometric metallicities 
deviate systematically from the spectroscopic values. The deviations are quite large for giants
as well as dwarfs at very low or high metallicities.
The bottom left panel of Fig.\,9 compares values of $\chi_{\rm min}^2 (SL_D)$ and $\chi_{\rm min}^2 (SL_G)$.
The majority of giants show a lower value of $\chi_{\rm min}^2 (SL_G)$ and the majority of dwarfs 
show a lower value of $\chi_{\rm min}^2 (SL_D)$, thus providing a way to effectively select giants against dwarfs.
Stars falling in the purple triangle region are red giant candidates thus selected. 
Their distributions in the \feh~(DR9) -- \logg~and  $g-i$ -- \logg~planes are shown  
in the middle and right bottom panels, respectively.
The selection efficiency is very high at low metallicities and insensitive to colors. 
Most contamination are from MS stars of \feh~between $-$1.0 -- $-$0.3.
In this metallicity, the stellar loci of MS stars of a given metallicity 
mimic those of red giants of a slightly different metallicity.

We divide the test sample into two-dimensional bins of $g-i$ and \feh.
Here the \feh~values refer to those from the SDSS DR9.
The ranges are 0.55 -- 1.15\,mag and $-$2.5 -- 0.0
and the steps 0.1\,mag and 0.1, respectively.
For each bin, the selection completeness, defined as the ratio of the number of red giants selected as 
candidates to the number of all red giants in the bin, and the selection efficiency, defined as 
the ratio of the number of red giants selected to the number of all red giant candidates in the bin,
are calculated. The two-dimensional distributions of selection completeness and efficiency as a function of $g-i$ and \feh~
are shown in Fig.\,10. The distributions integrated over color are plotted as a function of \feh~in Fig.\,11.
Note that some bins, mostly of \feh~ higher than $-0.3$ or $g-i$ redder than 1.05\,mag, 
have no red giants. For those bins, the completeness and efficiency are assigned to be zero. 
For the remaining bins, the efficiency shows a strong dependence on \feh~but only weakly on color. 
The efficiency is nearly 100 per cent for metal-poor giants of \feh~$\le -1.2$, decreasing slightly to 70 per cent for 
more metal-rich stars.
Above \feh~$\ge -1.2$, the efficiency decreases rapidly with \feh, from 60 per cent at $-$1.1, 
40 per cent at $-$1.0, to below 10 per cent at $-$0.5. 
The completeness shows some small variations, with a typical value of 70 per cent.
The above results show that the method proposed here is capable of selecting 
metal-poor red giants with a very high level of efficiency as well as completeness.  

To further test the efficiency and completeness of the method, we select a subsample of metal-poor stars 
of \feh~$\le -1.2$ and plot their HR diagram in Fig.\,12. The parallaxes are from the {\it Gaia} EDR3 \citep{2021A&A...649A...1G}. The black and purple dots represent candidates of MS stars and red giant stars, respectively. The red line in Fig.\,12 is used to separate red giant stars from dwarfs.
It can be seen that most candidates 
of red giant stars are above the red line, and most stars above the red line are purple dots. Quantitatively,  
under the new criterion, the method has efficiency and completeness of 63 per cent and 71 per cent, respectively.
The numbers are slightly smaller than previous results, due to different criterion of red giant stars adopted. 
Note that the efficiency and completeness of the method depend not only on the photometric errors of the data used 
but also on the spatial position (particularly the Galactic latitude) of the sample. The method is most suitable 
for high Galactic latitude regions.

The method is also tested using the same samples of red giant stars in the four GCs M\,92, M\,13, M\,3, and M\,5 described in the previous two Sections.
Completeness of 67, 72, 40, and 74 per cent is achieved for M\,92, M\,13, M\,3, and M\,5, respectively, 
consistent with the results above for field stars. 
Details of the results are illustrated in Fig.\,13. 
For M\,92, M\,13, and M\,5, there is a well-defined boundary in the  
plane of $\chi_{\rm min}^2$ (SL$_{\rm D}$) versus $\chi_{\rm min}^2$ (SL$_{\rm G}$).
If we loose the criterion slightly, then almost all the giants will be selected.
With the current criterion, i.e., $\chi_{\rm min}^2$ (SL$_{\rm D}$) $>$ $\chi_{\rm min}^2$ (SL$_{\rm G}$),  
those missed are mostly outliers in the color -- color diagrams.
The spatial distributions of stars selected and those missed are different, suggesting 
spatially-dependent (calibration) errors.
For M\,3, stars missed distribute tightly in the color -- color diagrams, but in regions quite different from those selected, 
suggesting large systematic errors.
The spatial distributions of stars selected and those missed in M\,3 are also different.
Note that M\,3 was imaged by the SDSS with the 3rd and 4th columns of CCDs,  
for neither of those the non-linearity of detectors in the $z$ band has not been fully corrected for \citep{2015ApJ...799..133Y}.
This may partly explain the large deviations of stars relative to the stellar loci 
in the direction of bluer $g-i$ colors (fainter magnitudes) in the $(g-i)$ -- $(i-z)$ diagram.
With better calibration in the future, 
one can expect much higher completeness than possible for the moment, especially for M\,3.
For example, after correcting for the photometric zero-point differences between the photometry of \cite{2008ApJS..179..326A} and 
the SDSS ubercal calibration \citep{2013ApJ...763...65A},
the completeness increases to 89, 67, and 79 per cent for M\,92, M\,3, and M\,5, respectively,
but slightly decreases to 67 per cent for M\,13.

\begin{figure*}
\includegraphics[width=\textwidth]{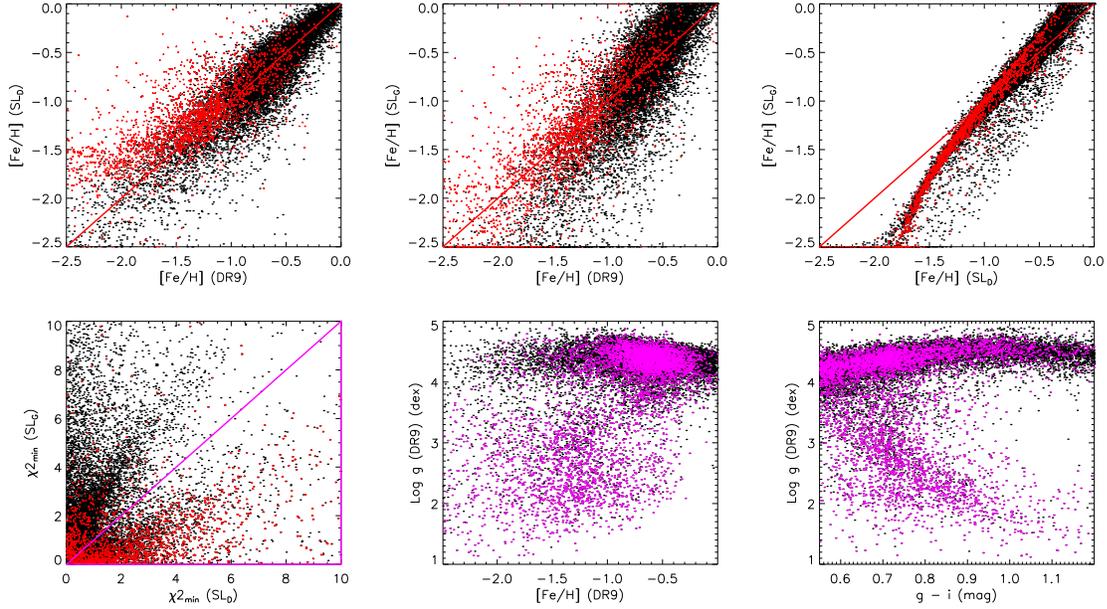}
\caption{
{\em Top panels:}\, Comparisons of spectroscopic metallicities from the SDSS DR9 with those estimated photometrically, 
using the stellar loci of dwarfs, \feh~(SL$_{\rm D}$), or using the loci of red giants, \feh~(SL$_{\rm G}$). 
The latter two sets of photometric estimates are also compared against the other.
The bottom left panel plots values of $\chi_{\rm min}^2 (SL_D)$ against $\chi_{\rm min}^2 (SL_G)$; 
In the above four panels, 
the black and red dots denote MS dwarfs and red giants in the test sample, respectively. The diagonal lines denote 
equal values of the two quantities in comparison.
The middle bottom panel plots \feh~against \logg~from SDSS DR9. The right  
bottom panel plots $g-i$ against \logg.
Black and purple dots in the bottom middle and right panels 
denote stars falling outside and inside the purple triangle delineated in the bottom left panel, respectively. 
}
\label{}
\end{figure*}

\begin{figure*}
\centering
\includegraphics[width=8cm]{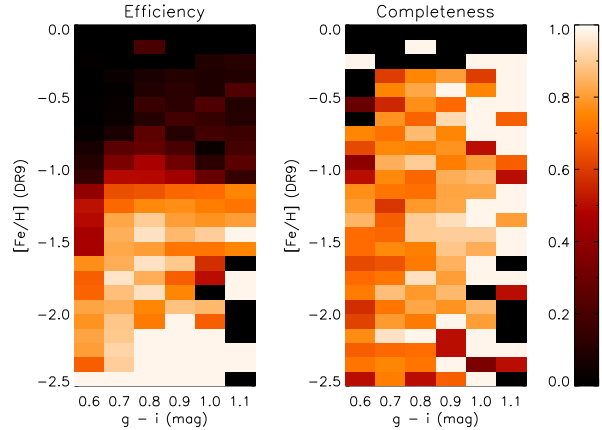}
\caption{Selection efficiency (left panel) and completeness (right panel) of red giants as a function of $g-i$ color and \feh.}
\label{}
\end{figure*}

\begin{figure*}
\centering
\includegraphics[width=8cm]{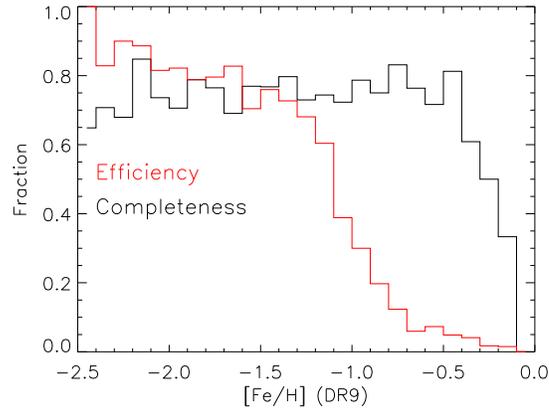}
\caption{Integrated completeness (black line) and efficiency (red line) as a function of \feh.}
\label{}
\end{figure*}

\begin{figure*}
\centering
\includegraphics[width=8cm]{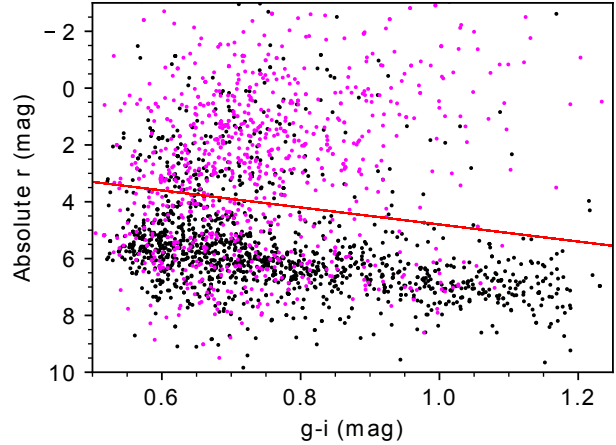}
\caption{HR diagram for stars of \feh~$\le -1.2$. The black and purple dots represent candidates of MS stars and red giant stars, respectively. The red line is used to separate red giant stars from dwarfs.}
\label{}
\end{figure*}
   
\begin{figure*}
\centering
\includegraphics[width=\textwidth]{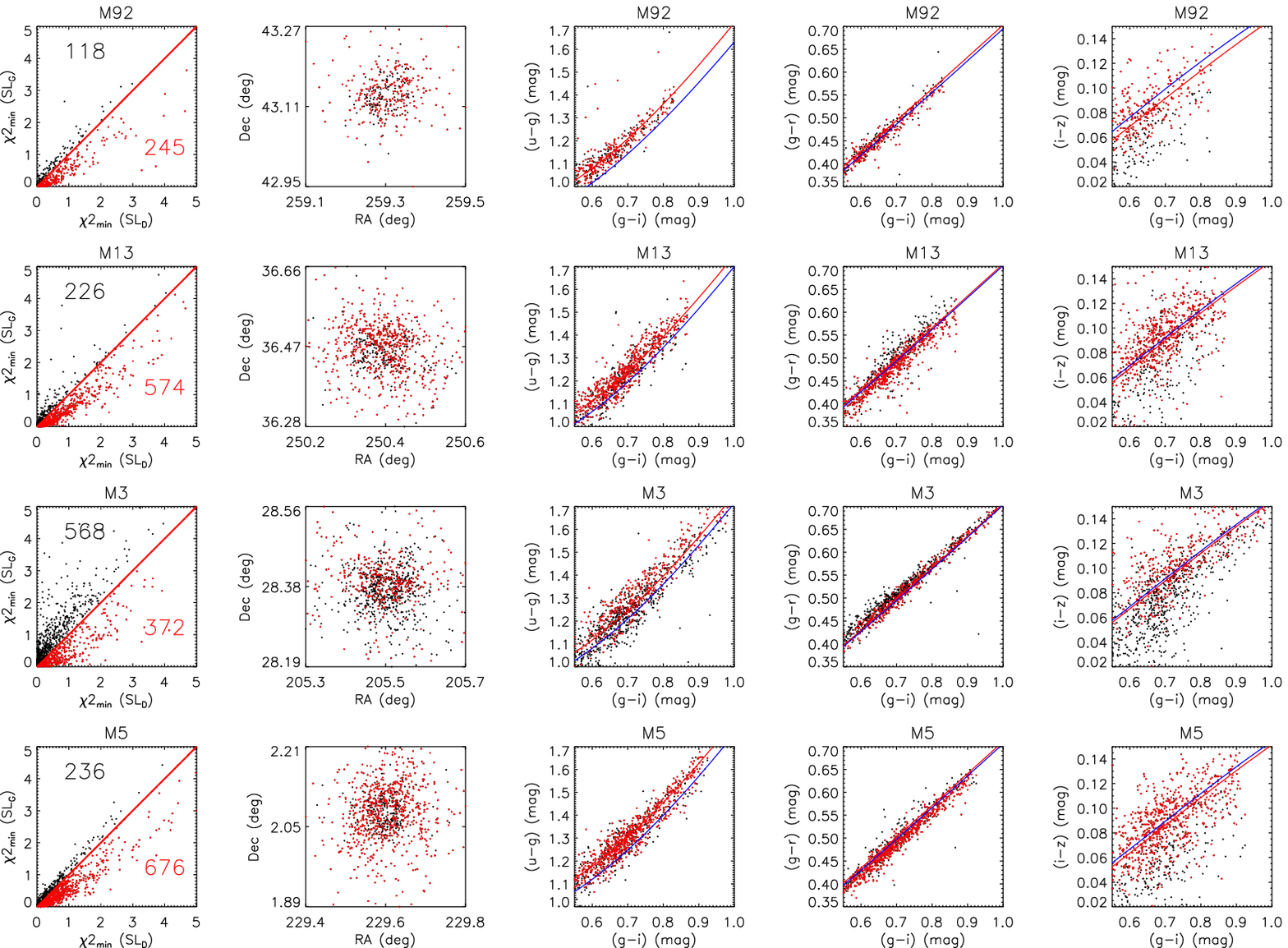}
\caption{
From left to right, the panels display distributions of the red giant stars in different GCs in the 
 $\chi_{\rm min}^2$ (SL$_{\rm D}$) versus $\chi_{\rm min}^2$ (SL$_{\rm G}$) ,  
RA versus Dec, $g-i$ versus $u-g$, $g-i$ versus $g-r$, and $g-i$ versus $i-z$ planes,  respectively.
The black and red dots denote candidates of MS stars and red giant stars, respectively.
Their numbers are labeled in black and red, respectively. 
The red and blue lines in the 3rd, 4th, and 5th columns denote the stellar loci at 
the average photometric metallicities and spectroscopic metallicities from the literature, respectively.
}
\label{}
\end{figure*}

\section{Summary}
In this paper,  by combining spectroscopic information and re-calibrated imaging photometry of 
the SDSS Stripe 82, we have built a sample of red giant stars with accurate colors and well-determined metallicities
to study the metallicity dependence and intrinsic widths of the SDSS stellar loci of red giant stars.
As in Paper\,I, we perform two-dimensional polynomial fits of colors $u-g$, $g-r$, $r-i$, and $i-z$ 
as a function of color $g-i$ and metallicity \feh.
We find that colors $u-g$, $g-r$, $r-i$, and $i-z$ of red giant stars can be 
accurately predicted by their $g-i$ colors and metallicities. 
The fit residuals, at the level of 0.032, 0.007, 0.007, and 0.011\,mag for $u-g$, $g-r$, $r-i$, and $i-z$, respectively,
are consistent with the photometric errors, metallicity determination uncertainties, and calibration errors, suggesting that the intrinsic widths
of loci of red giant stars are also at maximum a few mmag.
The results are further supported by analysis of red giant stars in four GCs, M\,92, M\,13, M\,3, and M\,5.

Systematic differences exist between the metallicity-dependent stellar loci of red giants and MS dwarfs.
The colors of giants are less sensitive to metallicity than MS dwarfs. 
The metallicity-dependent stellar loci of red giant stars can be used to estimate their photometric metallicities 
by simultaneously fitting the $u-g$, $g-r$, $r-i$, and $i-z$ colors.
A precision of 0.2 -- 0.25\,dex is achieved with the SDSS photometry, 
comparable to that achievable by low-resolution spectroscopy at SNR of 10. 
Tests with red giant stars in the four GCs  M\,92, M\,13, M\,3, and M\,5
show consistent results. The systematic discrepancies seen 
between the mean photometric and spectroscopic metallicities
are probably caused by the calibration errors in the photometric data used. 
It suggests that color calibration 
accurate to a few mmag, achievable with the SCR method of \cite{2015ApJ...799..133Y}, 
is preferred to obtain robust photometric metallicities, especially for very metal-poor red giant stars.  

Based on the systematic differences between the stellar loci of red giant stars and MS stars,
we have further proposed a new technique to discriminate red giant stars from MS stars
using the SDSS photometry only. 
The method achieves completeness of $\sim$ 70 per cent and efficiency of $\sim$ 80 per cent
in selecting metal-poor red giant stars of \feh~$\le -$1.2 with good photometry. 
Photometric metallicities of the selected candidates of red giant stars are yielded simultaneously.
With the technique, we expect to identify a large number of metal-poor red giant stars from in the Stripe\,82 region as well as other regions imaged by the SDSS, and future surveys including the Vera Rubin Observatory (LSST; \citealt{2009arXiv0912.0201L}) and the China Space Station Telescope (CSST; \citealt{2011SSPMA..41.1441Z}),
enabling us to probe the structure and assemblage history of the Galactic halo. 


\begin{acknowledgements}

This work is supported by the National Natural Science Foundation of China through the project NSFC 11603002,
the National Key Basic R\&D Program of China via 2019YFA0405503 and Beijing Normal University grant No. 310232102.
We acknowledge the science research grants from the China Manned Space Project with NO. CMS-CSST-2021-A08 and CMS-CSST-2021-A09.
   
Funding for SDSS-III has been provided by the Alfred P. Sloan Foundation, the Participating Institutions, the National Science Foundation, and the U.S.
Department of Energy Office of Science. The SDSS-III web site is http://www.sdss3.org/.SDSS-III is managed by the Astrophysical Research Consortium for the Participating Institutions of the SDSS-III Collaboration including the
University of Arizona, the Brazilian Participation Group, Brookhaven
NationalLaboratory, Carnegie Mellon University, University of Florida, the
French Participation Group, the German Participation Group, Harvard University, the
Instituto de Astrofisica de Canarias, the Michigan State/Notre Dame/JINA
Participation Group, Johns Hopkins University, Lawrence Berkeley National
Laboratory, Max Planck Institute for Astrophysics, Max Planck Institute for
Extraterrestrial Physics, New Mexico State University, New York University,
Ohio State University, Pennsylvania State University, University of Portsmouth,
Princeton University, the Spanish Participation Group, University of Tokyo,
University of Utah, Vanderbilt University, University of Virginia, University
of Washington, and Yale University.

\end{acknowledgements}

\bibliographystyle{raa}
\bibliography{bibtex}

\begin{thebibliography}{31}
\providecommand\natexlab[1]{#1}
\providecommand\JournalTitle[1]{#1}

\bibitem[{Ahn} {et~al.}(2012)]{2012ApJS..203...21A}
{Ahn}, C.~P., {Alexandroff}, R., {Allende Prieto}, C., {et~al.} 2012, \apjs,
  203, 21

\bibitem[{An} {et~al.}(2008)]{2008ApJS..179..326A}
{An}, D., {Johnson}, J.~A., {Clem}, J.~L., {et~al.} 2008, \apjs, 179, 326

\bibitem[{An} {et~al.}(2013)]{2013ApJ...763...65A}
{An}, D., {Beers}, T.~C., {Johnson}, J.~A., {et~al.} 2013, \apj, 763, 65

\bibitem[{Deng} {et~al.}(2012)]{2012RAA....12..735D}
{Deng}, L.-C., {Newberg}, H.~J., {Liu}, C., {et~al.} 2012, Research in
  Astronomy and Astrophysics, 12, 735

\bibitem[{Flynn} \& {Morrison}(1990)]{1990AJ....100.1181F}
{Flynn}, C., \& {Morrison}, H.~L. 1990, \aj, 100, 1181

\bibitem[{Gaia Collaboration} {et~al.}(2021)]{2021A&A...649A...1G}
{Gaia Collaboration}, {Brown}, A.~G.~A., {Vallenari}, A., {et~al.} 2021, \aap,
  649, A1

\bibitem[{Harris}(1996)]{1996AJ....112.1487H}
{Harris}, W.~E. 1996, \aj, 112, 1487

\bibitem[{Ivezi{\'c}} {et~al.}(2007)]{2007AJ....134..973I}
{Ivezi{\'c}}, {\v{Z}}., {Smith}, J.~A., {Miknaitis}, G., {et~al.} 2007, \aj,
  134, 973

\bibitem[{Kraft} \& {Ivans}(2004)]{2004oee..sympE..33K}
{Kraft}, R.~P., \& {Ivans}, I.~I. 2004, in Origin and Evolution of the
  Elements, ed. A.~{McWilliam} \& M.~{Rauch}, 33

\bibitem[{Lardo} {et~al.}(2011)]{2011A&A...525A.114L}
{Lardo}, C., {Bellazzini}, M., {Pancino}, E., {et~al.} 2011, \aap, 525, A114

\bibitem[{Lee} {et~al.}(2008{\natexlab{a}})]{2008AJ....136.2022L}
{Lee}, Y.~S., {Beers}, T.~C., {Sivarani}, T., {et~al.} 2008{\natexlab{a}}, \aj,
  136, 2022

\bibitem[{Lee} {et~al.}(2008{\natexlab{b}})]{2008AJ....136.2050L}
{Lee}, Y.~S., {Beers}, T.~C., {Sivarani}, T., {et~al.} 2008{\natexlab{b}}, \aj,
  136, 2050

\bibitem[{Liu} {et~al.}(2014)]{2014IAUS..298..310L}
{Liu}, X.~W., {Yuan}, H.~B., {Huo}, Z.~Y., {et~al.} 2014, in Setting the scene
  for Gaia and LAMOST, ed. S.~{Feltzing}, G.~{Zhao}, N.~A. {Walton}, \&
  P.~{Whitelock}, Vol. 298, 310

\bibitem[{LSST Science Collaboration} {et~al.}(2009)]{2009arXiv0912.0201L}
{LSST Science Collaboration}, {Abell}, P.~A., {Allison}, J., {et~al.} 2009,
  arXiv e-prints, arXiv:0912.0201

\bibitem[{Lupton} {et~al.}(2002)]{2002SPIE.4836..350L}
{Lupton}, R.~H., {Ivezic}, Z., {Gunn}, J.~E., {et~al.} 2002, in Society of
  Photo-Optical Instrumentation Engineers (SPIE) Conference Series, Vol. 4836,
  Survey and Other Telescope Technologies and Discoveries, ed. J.~A. {Tyson} \&
  S.~{Wolff}, 350

\bibitem[{Morrison} {et~al.}(2000)]{2000AJ....119.2254M}
{Morrison}, H.~L., {Mateo}, M., {Olszewski}, E.~W., {et~al.} 2000, \aj, 119,
  2254

\bibitem[{Padmanabhan} {et~al.}(2008)]{2008ApJ...674.1217P}
{Padmanabhan}, N., {Schlegel}, D.~J., {Finkbeiner}, D.~P., {et~al.} 2008, \apj,
  674, 1217

\bibitem[{Ratnatunga} \& {Freeman}(1985)]{1985ApJ...291..260R}
{Ratnatunga}, K.~U., \& {Freeman}, K.~C. 1985, \apj, 291, 260

\bibitem[{Schlegel} {et~al.}(1998)]{1998ApJ...500..525S}
{Schlegel}, D.~J., {Finkbeiner}, D.~P., \& {Davis}, M. 1998, \apj, 500, 525

\bibitem[{Skrutskie} {et~al.}(2006)]{2006AJ....131.1163S}
{Skrutskie}, M.~F., {Cutri}, R.~M., {Stiening}, R., {et~al.} 2006, \aj, 131,
  1163

\bibitem[{Stetson}(1987)]{1987PASP...99..191S}
{Stetson}, P.~B. 1987, \pasp, 99, 191

\bibitem[{Stetson}(1994)]{1994PASP..106..250S}
{Stetson}, P.~B. 1994, \pasp, 106, 250

\bibitem[{Wright} {et~al.}(2010)]{2010AJ....140.1868W}
{Wright}, E.~L., {Eisenhardt}, P. R.~M., {Mainzer}, A.~K., {et~al.} 2010, \aj,
  140, 1868

\bibitem[{Yanny} {et~al.}(2009)]{2009AJ....137.4377Y}
{Yanny}, B., {Rockosi}, C., {Newberg}, H.~J., {et~al.} 2009, \aj, 137, 4377

\bibitem[{York} {et~al.}(2000)]{2000AJ....120.1579Y}
{York}, D.~G., {Adelman}, J., {Anderson}, John~E., J., {et~al.} 2000, \aj, 120,
  1579

\bibitem[{Yuan} {et~al.}(2013)]{2013MNRAS.430.2188Y}
{Yuan}, H.~B., {Liu}, X.~W., \& {Xiang}, M.~S. 2013, \mnras, 430, 2188

\bibitem[{Yuan} {et~al.}(2015{\natexlab{a}})]{2015ApJ...799..134Y}
{Yuan}, H., {Liu}, X., {Xiang}, M., {Huang}, Y., \& {Chen}, B.
  2015{\natexlab{a}}, \apj, 799, 134

\bibitem[{Yuan} {et~al.}(2015{\natexlab{b}})]{2015ApJ...803...13Y}
{Yuan}, H., {Liu}, X., {Xiang}, M., {Huang}, Y., \& {Chen}, B.
  2015{\natexlab{b}}, \apj, 803, 13

\bibitem[{Yuan} {et~al.}(2015{\natexlab{c}})]{2015ApJ...799..135Y}
{Yuan}, H., {Liu}, X., {Xiang}, M., {et~al.} 2015{\natexlab{c}}, \apj, 799, 135

\bibitem[{Yuan} {et~al.}(2015{\natexlab{d}})]{2015ApJ...799..133Y}
{Yuan}, H., {Liu}, X., {Xiang}, M., {et~al.} 2015{\natexlab{d}}, \apj, 799, 133

\bibitem[{Zhan}(2011)]{2011SSPMA..41.1441Z}
{Zhan}, H. 2011, Scientia Sinica Physica, Mechanica \& Astronomica, 41, 1441

\end{thebibliography}

\label{lastpage}

\end{document}